\documentclass{aa}
\usepackage{graphics}
\usepackage{epsfig}
\usepackage{times,mathptm}
\def\tabvd{\mbox{$\stackrel{\displaystyle .}{\displaystyle .}$}}
\begin{document}

\thesaurus{05         
           (08.05.1;  
            08.08.1;  
            08.06.2;  
            09.02.1;  
            09.09.1 LMC$\,$4;
            11.13.1)  
            }         
   \title{The stars near the centre of supergiant shell LMC$\,$4: \\
          Further constraints on triggering scenarios\thanks{Based on
          observations collected at the European Southern Observatory (ESO),
          La Silla, Chile.}
          }

   \titlerunning{The centre of SGS LMC$\,$4: Further constraints
                 on triggering scenarios}

   \author{Jochen M. Braun \and  Klaas S. de Boer \and  Martin Altmann
          }

   \authorrunning{J.M. Braun et al.}

   \offprints{
{\tt 'jbraun@{\discretionary{}{}{}}astro.{\discretionary{}{}{}}uni-bonn{\discretionary{}{}{}}.de'}}

   \institute{Sternwarte der Universit\"at Bonn, Auf dem H\"ugel 71,
              D--53121 Bonn, Germany\\
              WWW-URL: http://www.astro.uni-bonn/$\,\,\tilde{}\,\,$webstw
             }

   \date{Received June, 2000; accepted Month, Year}

   \maketitle

   \begin{abstract}

The huge supergiant shell (SGS) LMC$\,$4, next to the giant H$\,${\sc ii}
region 30$\,$Doradus the second most impressive feature of the Large Magellanic
Cloud (LMC), has been the subject of many studies and was used to test various
models of star forming mechanisms.

In this paper we present a $B,V$ photometry of the very centre of this SGS.
The new data yield an age of $11\,(2)\;$Myr and a $B-V$ excess of
$0.10\,(3)\;$mag.
Additionally, the two clusters near the geometric centre are too old to be
related to the formation of LMC$\,$4:
HS$\,$343 being $\sim 0.1\;$Gyr and KMHK$\,$1000 being $\sim 0.3\;$Gyr of age.
This and existing photometries support that the young stellar population
covering the entire inner region is nearly coeval.

We compare these findings with predictions of models proposed for
the creation mechanism of SGSs.
We conclude that
the large-scale trigger, necessary to explain the observations,
comes from hydrodynamic interaction of the galactic halo with gas in
the leading edge of the moving LMC.

   \end{abstract}


\section{Introduction}
\label{s_intro}

The creation mechanism for interstellar shells of diameters larger than 300~pc
continues to be debated.
Such sizes rule out that a single (central) stellar association can provide
the energy (the combined effect of SN explosions,
stellar winds and radiation pressure) to create such a supershell
(see Tenorio-Tagle \& Bodenheimer 1988 for a review).
Apparently much more energy is needed so that different formation mechanisms
have been considered.

In the Large Magellanic Cloud (LMC) Meaburn and collaborators
(see e.g. Meaburn 1980) have identified nine supergiant shells (SGSs)
based on their filamentary ring-like structure on H$\alpha$ plates.
These SGSs have diameters in the range of $0.6-1.4\;$kpc.
The largest and best studied is LMC$\,$4,
which appears in the light of H$\alpha$ as an ellipse of 1~kpc (in E-W) and
1.4~kpc diameter (in N-S direction).
In H$\,${\sc i} it is a thick shell with a gas depleted central area
(see, e.g., Luks \& Rohlfs 1992; Kim et al. 1999).

The inner part of LMC$\,$4 is dominated by the stellar superassociation
LH$\,$77.
In literature one finds the name Shapley Constellation III, which is
also used for the entire inner part of this SGS.
Recently Efremov \& Elmegreen (1998) pointed out that this term was used
for the wrong object.
Originally McKibben Nail \& Shapley (1953) `defined' it to be
a triple aggregation of $26' \times 26'$ area and to contain NGC$\,$1974.
So this name would not be appropriate for the LH$\,$77 region
(see Fig.~\ref{f_lmcsgslhsketch}),
but for the N$\,$51 region,
containing LH$\,$63, LH$\,$60, and LH$\,$54.
Thus we consider it best to avoid the ambiguous term Sh$\;$III and to only
use appropriate object names based on a clear naming convention,
like LH$\,nn$ (Lucke~\& Hodge 1970; Lucke 1972), N$\,nn$ (Henize 1956), or
LMC$\,$4 (Goudis~\& Meaburn 1978).

\begin{figure}
\epsfxsize=8.4cm
\centerline{\epsffile{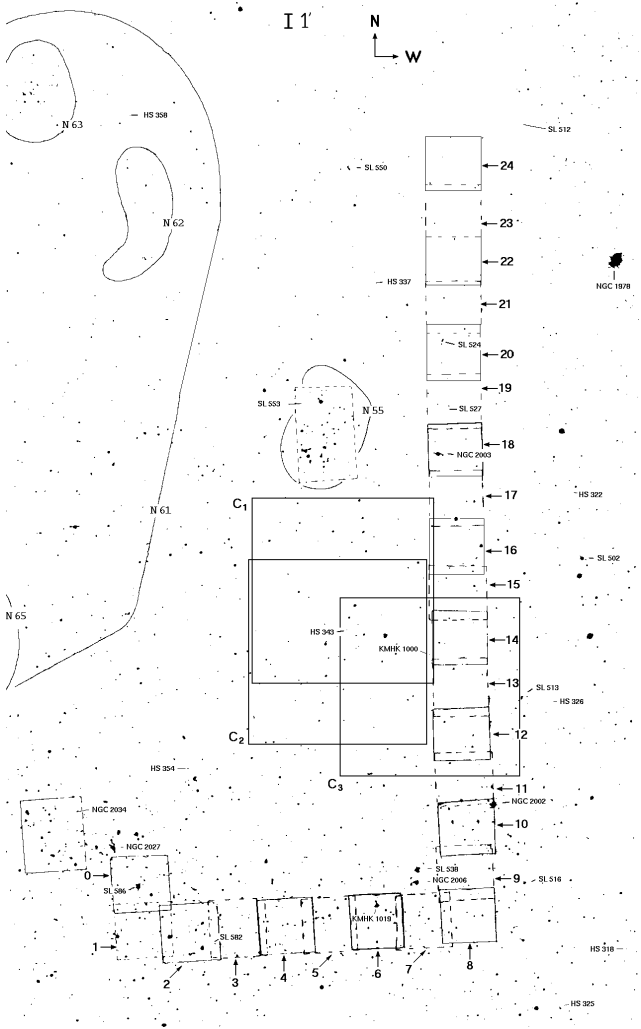}}
\caption[]{Mosaic of
the central region of LMC$\,$4 out of four $V$ charts (44, 45, 51, 52)
of the LMC atlas of Hodge \& Wright (1967).
The 25 fields of the 'J'-shaped area (Braun et al. 1997) and
the 3~centre fields C$_{1-3}$ (see Sect.~\ref{s_lmc4c_obsred})
are outlined}
\label{f_lmc4c_hw}
\end{figure}

Since the identification of LMC$\,$4 as SGS,
several photometric investigations have been carried out.
These aimed, in part, at finding the age of the stars inside LMC$\,$4
and at its rim with the goal of tracing the history of the SGS.
These studies include those of
Isserstedt (1984) and Reid et al. (1987),
which indicated propagating star fomation,
and the recent CCD photometries of
Braun et al. (1997) and Dolphin~\& Hunter (1998),
showing a nearly coeval population without any age gradient
(see Sect.~\ref{s_photdat}).
By now, the CCD studies, in combination with new data presented in Sect.~2,
cover about half of the area inside the H$\alpha$ rim.
All these studies of the interior agree on the young age of the stars,
being of the order of 11-13~Myr,
with very little spread.
Of course, a much older background population is present, too.
But it must be the young stars which are related with the formation
of the SGS LMC$\,$4.

Additional studies of star groups near the outer edge include those about
NGC$\,$1948 (Vallenari et al. 1993; Will et al. 1996),
LH$\,$63, LH$\,$60, LH$\,$54 (Petr et al. 1994),
LH$\,$76 (Wilcots et al. 1996),
LH$\,$95, LH$\,$91 (Gouliermis et al. 2000),
NGC$\,$2030 (Laval et al. 1986),
and the association LH$\,$72 (Olsen et al. 1997)
in the (2D projected) central region.
The surprising result of these studies is,
that the stars {\it in} the SGS (those near the H$\alpha$ emission)
have an age only marginally younger than those in the interior.

The small difference in age between the interior and the edge of LMC$\,$4
makes it likely that the stars near the H$\alpha$ emission were formed
as a result of the events taking place inside the SGS.
However, the sheer volume of space interior to LMC$\,$4
containing stars of the same age makes LMC$\,$4 a truly exceptional area.

In the present study we have compiled the data and the results
on ages and population structure for the LMC$\,$4 area.
We also add some new photometry, consisting of three fields
near the centre covering a total of $167.3\;\sq '$.

The overall data are analysed to find further clues
to the origin of the SGS LMC$\,$4.
Various mechanisms have been proposed for the formation of LMC$\,$4.
These include
the collision with a high-velocity H$\,${\sc i} cloud (HVC),
the stochastic self-propagating star formation (SSPSF),
the triggering of star cluster arcs,
and the LMC bow-shock as star formation trigger.
These will be discussed,
including remarks about their viability and verifiability.

\section{Observations and data reduction}
\label{s_lmc4c_obsred}

The new data\footnote{
  The entire Table~\ref{t_lmc4cphot} of this publication is only available
  electronically, at the CDS (see Editorial in A\&A 280, E1, 1993) or at
  the Astronomical Institutes of Bonn University
  ({\tt 'ftp ftp.{\discretionary{}{}{}}astro.{\discretionary{}{}{}}uni-bonn.{\discretionary{}{}{}}de'};
   further information can be obtained
   at the URL
   {\tt 'http://{\discretionary{}{}{}}www.{\discretionary{}{}{}}astro.{\discretionary{}{}{}}uni-bonn.{\discretionary{}{}{}}de/{\discretionary{}{}{}}{$\,\,\tilde{}\,\,$}jbraun/{\discretionary{}{}{}}phdt\_lmc4c.{\discretionary{}{}{}}html'}).}
were taken in January, 1999 with the $1.54\;$m
Danish Telescope at ESO observatory on Cerro La Silla.
The telescope was equipped with DFOSC and the 2k$^2\;$pix$^2$ LORAL CCD
(W7 Chip) resulting in a scale factor of \mbox{$0.39''$ pix$^{-1}$}.
The chip possesses one 6-fold, six double and six single bad columns plus a
5-fold bad line produced by the electronics on two object frames.
Additionally, there were some flatfield problems causing a residual large-scale
gradient of up to 4\% in $V$ sky flats, so flatfields had to be derived by
combining object frames and twilight flats.

The data reduction was performed on GNU/Linux (i386) workstations with MIDAS
(see e.g. Banse et al. 1983), IRAF (Tody 1986)
and DAOPHOT~II (Stetson 1987).

The calibration was achieved using the standard fields SA$\,95-42$,
Rubin$\,149$, PG$\,1047+003$, and PG$\,0918+029$ (Landolt 1992), observed
in $B,V$ passbands at different airmasses on 14th January 1999,
yielding an atmospheric extinction of $k_V = 0.163\,(32)\;$mag and
$k_B = 0.297\,(42)\;$mag and the calibration constants of
Eqs.~(\ref{e_lmc4c_calbv}) and (\ref{e_lmc4c_calv}).
After normalization (indicated by the index 'n') to airmass 0 and 1~s exposure
time including the PSF to aperture shift ($\delta_V = -0.066\,(15)\;$mag and
$\delta_B = -0.005\,(20)\;$mag), the following relations were applied:
\begin{equation}
\label{e_lmc4c_calbv}
(B - V) =
\begin{array}[t]{l}
  \left[ (B - V)_{\rm n} - 0.261\,(46)\;\mbox{mag} \right] \, / \,
  0.931\,(18)
\end{array}
\end{equation}
\begin{equation}
\label{e_lmc4c_calv}
V =
\begin{array}[t]{l}
  V_{\rm n} - 0.683\,(15)\;\mbox{mag} + 0.021\,(6) \cdot (B - V) 
\end{array}
\end{equation}
%

\begin{table}
\caption[]{Mean DAOPHOT errors and standard deviations of stellar
$V$~magnitudes and $B-V$ colours and number of stars in a given magnitude
range}
\label{t_lmc4cphoterr}
\begin{tabular}{cc@{\ }c@{\ }c@{\ }c@{\ }ccrll}
\hline
\noalign{\smallskip}
\multicolumn{7}{c}{Range} & \multicolumn{1}{c}{Number} &
 \multicolumn{1}{c}{$\overline{\Delta V}$} &
 \multicolumn{1}{c}{$\overline{\Delta (B-V)}$} \\
\multicolumn{7}{c}{[mag]} & \multicolumn{1}{c}{of stars} &
 \multicolumn{1}{c}{[mag]} & \multicolumn{1}{c}{[mag]} \\
\noalign{\smallskip}
\hline
\noalign{\smallskip}
 & & & $V$ & $<$ & 18 & & $2\,240$ & $0.019\,(15)$ & $0.031\,(22)$ \\
 & 18 & $\le$ & $V$ & $<$ & 20 & & $11\,224$ & $0.025\,(16)$ & $0.041\,(26)$ \\
 & 20 & $\le$ & $V$ & & & & $33\,285$ & $0.071\,(51)$ & $0.127\,(90)$ \\
\noalign{\smallskip}
\hline
\end{tabular}
\end{table}

\begin{figure*}
\epsfig{file=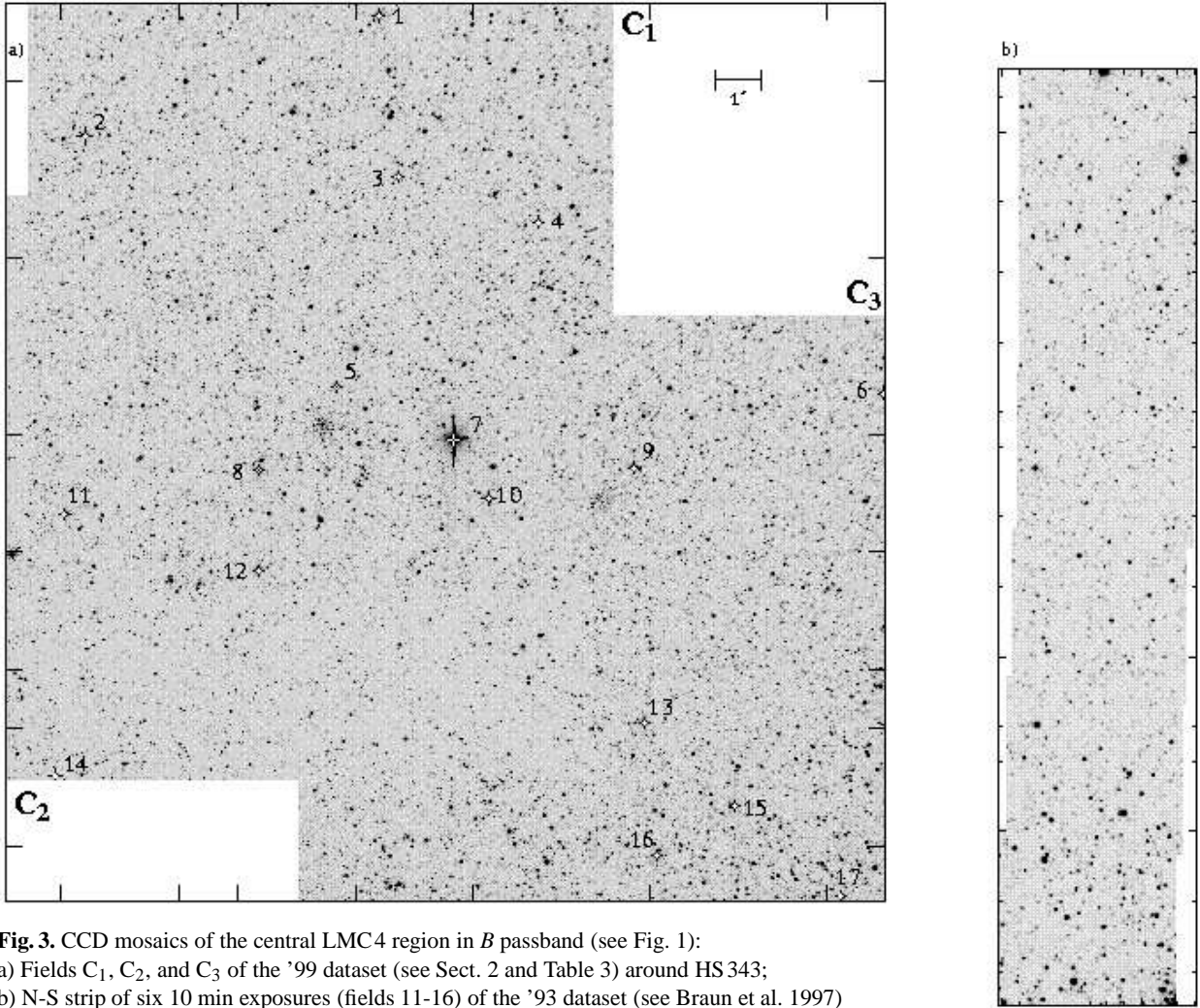,width=16.45cm,angle=0,
        bbllx=35.0, bblly=201.0, bburx=558.0, bbury=641.0, clip=}
\vspace*{-1.4cm}
\caption[]{CCD mosaics of the central LMC$\,$4 region in $B$ passband
(see Fig.~\ref{f_lmc4c_hw}): \\
a) Fields C$_1$, C$_2$, and C$_3$ of the '99 dataset (see
Sect.~\ref{s_lmc4c_obsred} and Table~\ref{t_lmc4cskmacs}) around HS$\,$343; \\
b) N-S strip of six $10\;$min exposures (fields 11-16) of the '93 dataset
(see Braun et al. 1997)}
\label{f_lmc4jstrip}\label{f_lmc4cmos}
\end{figure*}

The dataset contains the three fields C$_{1-3}$ of $167.3\;\sq '$ area
(see Fig.~\ref{f_lmc4c_hw} for their location and Fig.~\ref{f_lmc4cmos}a
for an image composit).
Each was observed in $B,V$ passbands with short (30 and $15\;$s) and long
(480 and $240\;$s, respectively) exposures (for field C$_3$ the exposure times
were 240 and $480\;$s) during a typical seeing of $1.5{''}$.
The entire CCD mosaic shown in Fig.~\ref{f_lmc4cmos}a covers $328.5\;\sq '$.

\begin{figure*}
\resizebox{\hsize}{!}{\includegraphics{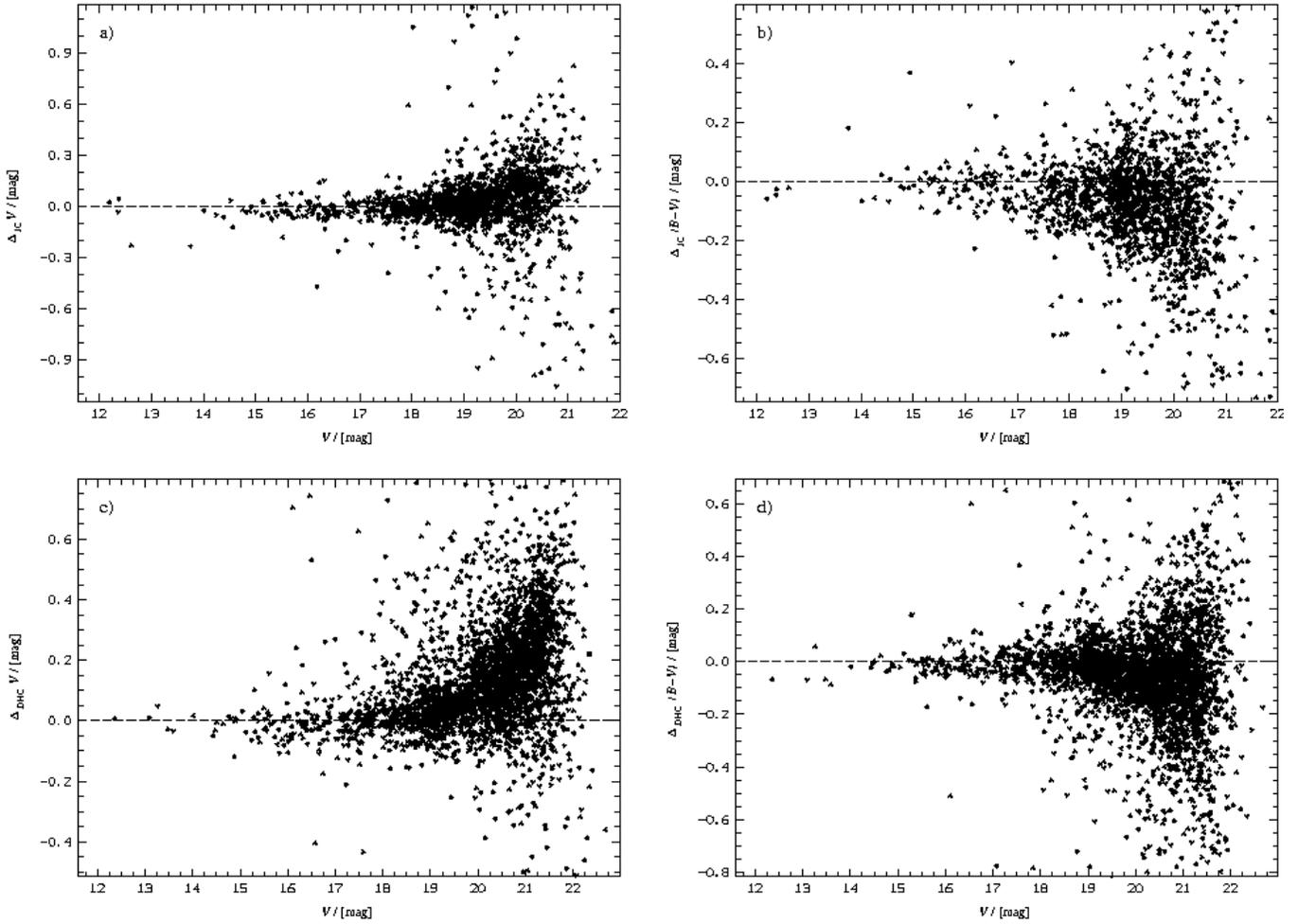}}
\caption[]{Photometric comparison of the field C$_3$ with the overlapping
regions of the 'J' dataset (a and b; see Braun et al. 1997) and
with the LH$\,$77 field (c and d; see Dolphin \& Hunter 1998).
The~zero difference lines are indicated.
For details see end of Sect.~\ref{s_lmc4c_obsred} and Table~\ref{t_lmc4c2j}
\vspace*{-0.25cm} \hspace*{17.5cm}}
\label{f_lmc4c2j}
\end{figure*}

\begin{table}
\caption[]{Statistics of the comparison of the three big photometries
inside LMC$\,$4.
The mean (with deviation) of the magnitude and the colour difference 
is given (compare with Fig.~\ref{f_lmc4c2j})
}
\label{t_lmc4c2j}
\begin{tabular}{l@{$\;\;\;$}r@{$\;\;$}cr@{$\;\;$}cc}
\hline
\noalign{\smallskip}
 Fig. &
 \multicolumn{1}{c@{$\;\;$}}{$\left(\overline{\Delta m}\right)_{\rm br}$} &
 \multicolumn{1}{@{$\;\;$}c}{$N_{\rm br}\;\:$} &
 \multicolumn{1}{c}{$\left(\overline{\Delta m}\right)_{\rm all}$} & \multicolumn{1}{c}{$N_{\rm all}$} &
 Range of $\Delta m$ \\
 & \multicolumn{1}{c}{[mag]} & & \multicolumn{1}{c}{[mag]} & &
 \multicolumn{1}{c}{[mag]} \\
\noalign{\smallskip}
\hline
 \ref{f_lmc4c2j}a &
 $-0.019\,(82)$ & 243 &  $0.015\,(387)$ & 1739 & $[\,-3.71\,,\,5.04\,]$ \\
 \ref{f_lmc4c2j}b &
 $-0.033\,(74)$ & 239 & $-0.076\,(223)$ & {''} & $[\,-1.46\,,\,1.78\,]$ \\
 \ref{f_lmc4c2j}c &
 $0.007\,(98)$ & 244 &  $0.140\,(219)$ & 3174 & $[\,-1.24\,,\,4.43\,]$ \\
 \ref{f_lmc4c2j}d &
 $-0.008\,(57)$ & 239 & $-0.063\,(251)$ & {''} & $[\,-4.87\,,\,2.97\,]$ \\
\hline
\noalign{\smallskip}
\multicolumn{6}{l}{
\begin{minipage}[t]{8.3cm}
 \ref{f_lmc4c2j}a: 
    $\Delta_{\rm JC} V := V_{\rm J,11-15}-V_{\rm C_3}$\\
 \ref{f_lmc4c2j}b: 
    $\Delta_{\rm JC} (B-V) := (B-V)_{\rm J,11-15}-(B-V)_{\rm C_3}$\\
 \ref{f_lmc4c2j}c: 
    $\Delta_{\rm DHC} V := V_{\rm DH,LH77}-V_{\rm C_3}$\\
 \ref{f_lmc4c2j}d: 
    $\Delta_{\rm DHC} (B-V) := (B-V)_{\rm DH,LH77}-(B-V)_{\rm C_3}$\\
The brighter subset (indicated by `br') is defined by \\
 \ref{f_lmc4c2j}a and \ref{f_lmc4c2j}c: 
 $V < 18\;$mag$\;$ and $\;\Delta V \in [-0.6,0.6]\;$mag$\;\:$ or \\
 \ref{f_lmc4c2j}b and \ref{f_lmc4c2j}d: 
 $V < 18\;$mag$\;$ and $\;\Delta (B-V) \in [-0.3,0.25]\;$mag
\end{minipage} }
\end{tabular}
\end{table}

The brightest star located near the centre of the analyzed area
is the galactic F8 dwarf HD 37195 \mbox{($\equiv$ P~1313} 
\mbox{$\equiv$ MACS~J0532$-$666\#013}, see Table~\ref{t_lmc4cskmacs}, \#7).
The compact cluster HS$\,$343 = KMHK$\,$1030 (Hodge \& Sexton 1966;
Kontizas et al. 1990) of $\sim 15\;$pc diameter can be found 2.9~arcmin to
the east of star \#7 while 2.5~arcmin to the west of star \#10
(Table~\ref{t_lmc4cskmacs} and Fig.~\ref{f_lmc4cmos}a) the loose cluster
KMHK$\,$1000 of $\sim 13\;$pc diameter is located.

As the western part of the centre field C$_3$ overlaps with the N-S strip
of the 'J'-shaped region (see Fig.~\ref{f_lmc4c_hw}),
Fig.~\ref{f_lmc4jstrip} shows both CCD mosaics at the same scale, so
e.g. stars marked \#9, 13, 15, and 16 on the left panel can easily be found
in the right panel.


\begin{table*}
\caption[]{Cross identification of the stars in the central LMC$\,$4 area
with catalogues by Sanduleak (1969; Sk) or Fehrenbach et al. (1970; P$|$G),
spectroscopic and photometric results compiled by Rousseau et al. (1978;
except values in '),
IDs in the Magellanic Catalogue of Stars (MACS; Tucholke et al. 1996),
and the results of our photometry.
The selected stars are labeled in Fig.~\ref{f_lmc4cmos}}
\label{t_lmc4cskmacs}
\begin{tabular}{rc@{$\;\;\;\:$}c@{$\;$}c@{$\;\;\:$}r@{$\;\;\:$}r@{$\;\;\:$}rc@{$\;\;\;$}r@{ }r@{ }r@{}l@{$\;\;\:$}r@{ }r@{ }r@{}l@{$\;\;\;\,$}r@{$\;\;\:$}r}
\hline
\noalign{\smallskip}
No. & Name & Sp & LC &
 \multicolumn{1}{c}{$\!\!\!V$} & $B-V$ & $U-B$ & MACS &
 \multicolumn{4}{c}{$\alpha$} & \multicolumn{4}{c}{$\delta$} &
 \multicolumn{1}{c}{$V$} & $B-V$ \\ 
\# & & & & [mag] & [mag] & [mag] & Name  & [$^{\rm h}$ & $^{\rm m}$
   & $^{\rm s}$ &] & [$^\circ$ & $'$ & $''$ &] & [mag] & [mag] \\
\noalign{\smallskip}
\hline
 1 & Sk $-66$ 128 & B0 & -- & 12.71 & $-0.12$ & $-1.01$ &
     J0531$-$665\#036 & 5 & 31 & 53&.490 & $-66$ & 31 & 14&.62 &
     12.83 & $-0.13$ \\ 
 2 & \multicolumn{1}{c}{--} & -- & -- & \multicolumn{1}{c}{--$\;\;$} &
     \multicolumn{1}{c}{--} & \multicolumn{1}{c}{--} &
     J0532$-$665\#041 & 5 & 32 & 59&.205 & $-66$ & 33 & 49&.96 &
     12.49 & $0.97$ \\ 
 3 & \multicolumn{1}{c}{--} & -- & -- & \multicolumn{1}{c}{--$\;\;$} &
     \multicolumn{1}{c}{--} & \multicolumn{1}{c}{--} &
     J0531$-$665\#033 & 5 & 31 & 49&.591 & $-66$ & 34 & 49&.82 &
     12.28 & $0.45$ \\ 
 4 & Sk $-66$ 121 & B1 & -- & 13.50 & $-0.19$ & $-1.05$ &
     J0531$-$665\#011 & 5 & 31 & 18&.150 & $-66$ & 35 & 51&.18 &
     13.53 & $-0.16$ \\ 
 5 & G 359 & A9 & I & 12.51 & 0.24 & 0.23 &
     J0532$-$666\#002 & 5 & 32 & 3&.736 & $-66$ & 39 & 28&.62 &
     12.54 & $0.19$ \\ 
 6 & Sk $-66$ 111 & B3 & I & 13.62 & $-0.17$ & $-0.88$ &
     J0530$-$666\#001 & 5 & 30 & 1&.331 & $-66$ & 39 & 43&.56 &
     \multicolumn{1}{c}{--} & \multicolumn{1}{c}{--} \\
 7 & P 1313 & 'F8$\,\,$ & $\,\,$V' &
     '9.38'$\!\!\!$ & '0.52'$\!\!\!$ & \multicolumn{1}{c}{--} &
     J0531$-$666\#013 & 5 & 31 & 37&.185 & $-66$ & 40 & 40&.57 &
     \multicolumn{1}{c}{--} & \multicolumn{1}{c}{--} \\ 
 8 & Sk $-66$ 130 & A0 & I & 12.05 & $0.04$ & $-0.37$ &
     \multicolumn{1}{c}{--} & 5 & 32 & 21&.212 & $-66$ & 41 & 18&.65 &
     12.06 & $0.07$ \\ 
 9 & Sk $-66$ 120 & A0 & I & 12.50 & $0.03$ & $-0.31$ &
     J0530$-$666\#039 & 5 & 30 & 56&.998 & $-66$ & 41 & 19&.96 &
     12.55 & $0.04$ \\ 
10 & Sk $-66$ 124 & A1 & I & 12.46 & $0.04$ & $-0.23$ &
     J0531$-$666\#012 & 5 & 31 & 29&.605 & $-66$ & 41 & 59&.23 &
     12.48 & $0.06$ \\ 
11 & G 375 & B6 & I & '13.10'$\!\!\!$ & \multicolumn{1}{c}{--} &
     \multicolumn{1}{c}{--} &
     J0533$-$667\#004 & 5 & 33 & 4&.862 & $-66$ & 42 & 14&.74 &
     12.98 & $-0.03$ \\ 
12 & Sk $-66$ 129 & A0 & I & 12.16 & $0.06$ & $-0.19$ &
     J0532$-$667\#015 & 5 & 32 & 21&.704 & $-66$ & 43 & 33&.72 &
     12.15 & $0.07$ \\ 
13 & Sk $-66$ 119a$\!\!\!$ & A0 & I & 12.16 & $0.07$ & $0.00$ &
     J0530$-$667\#030 & 5 & 30 & 55&.034 & $-66$ & 47 & 1&.30 &
     12.19 & $0.11$ \\ 
14 & \multicolumn{1}{c}{--} & -- & -- &
     \multicolumn{1}{c}{--$\;\;$} &
     \multicolumn{1}{c}{--} & \multicolumn{1}{c}{--} &
     J0533$-$668\#006 & 5 & 33 & 7&.603 & $-66$ & 48 & 5&.55 &
     12.50 & $1.82$ \\ 
15 & Sk $-66$ 117 & B2 & -- & 12.36 & $-0.06$ & $-0.76$ &
     J0530$-$668\#029 & 5 & 30 & 35&.007 & $-66$ & 48 & 53&.05 &
     12.36 & $-0.03$ \\ 
16 & Sk $-66$ 119 & B8 & I & 12.17 & $0.16$ & $-0.34$ &
     J0530$-$668\#051 & 5 & 30 & 52&.181 & $-66$ & 49 & 58&.49 &
     12.37 & $0.03$ \\ 
17 & \multicolumn{1}{c}{--} & -- & -- & '13.45'$\!\!\!$ &
     '$-0.14$'$\!\!\!$ & '$-0.93$'$\!\!\!$ &
     J0530$-$668\#010 & 5 & 30 & 10&.589 & $-66$ & 50 & 54&.92 &
     13.47 & $-0.08$ \\ 
\hline
\end{tabular}
\end{table*}

\begin{figure*}
\epsfxsize=18.96cm
\hspace*{-0.75cm}\epsffile{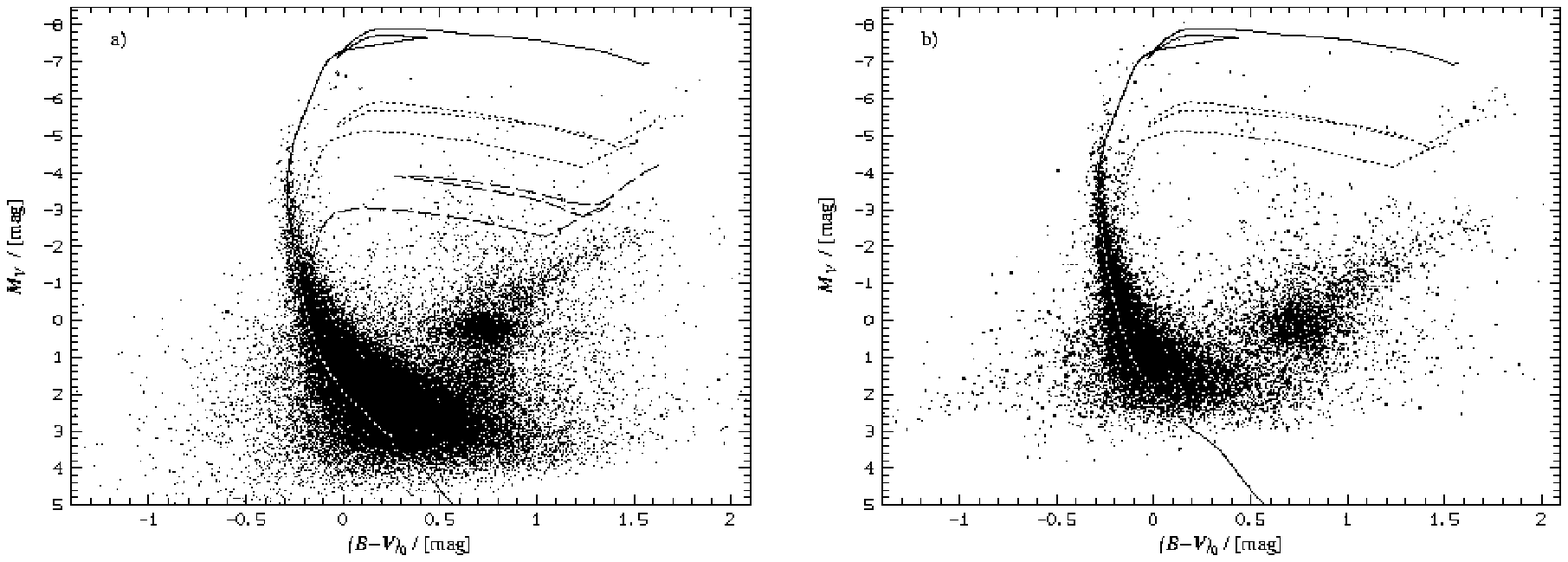}
\caption[]{CMDs of LMC$\,$4 with the isochrones of the Geneva group (Schaerer
et al. 1993) for LMC metallicity ($Z = 0.008$) and for logarithmic ages
$\;\log\,(t/[{\rm yr}]) \in \{ 7.05, 7.5, 8.0 \}$ and extinction correction
of $A_V = 0.31\;$mag.\\
a) CMD with $46\,749$ data points of the LMC$\,$4 centre fields C$_1$, C$_2$,
   and C$_3$ [left panel].\\
b) CMD with $15\,787$ data points of the LMC$\,$4 'J' dataset (Braun et al.
   1997) without fields 11-16 [right panel].\\
In panel a) the crosses near $M_V \approx -6.5\;$mag are 6 A supergiants
(see end of Sect.~\ref{s_photdat})
}
\label{f_lmc4ccmds}
\end{figure*}

\begin{figure*}
\epsfxsize=18.96cm
\hspace*{-0.75cm}\epsffile{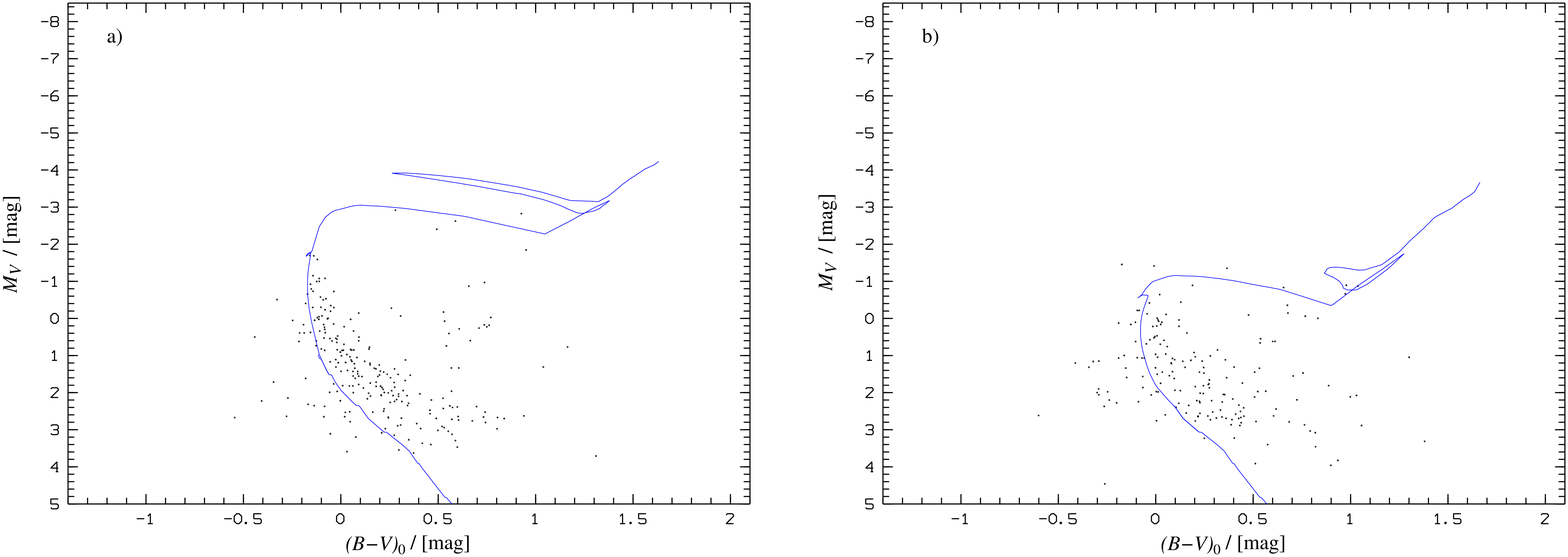}
\caption[]{CMDs of two clusters in the central region of LMC$\,$4, i.e., part
of the data plotted in Fig.~\ref{f_lmc4ccmds}a.\\
a) CMD of the circular region with radius of $34{''} = 8.3\;$pc around
   HS$\,$343 containing 237 data points [left panel].\\
b) CMD of the circular region with radius of $30{''} = 7.3\;$pc around
   KMHK$\,$1000 with 175 data points [right panel].\\
The isochrones of the Geneva group (Schaerer et al. 1993) of $0.1\;$Gyr
and $0.3\;$Gyr show that these stars are 'old' and are not related in any
way with events leading to the existence of supergiant shell LMC$\,$4
\vspace*{0.35cm}}
\label{f_lmc4cCLcmds}
\end{figure*}

\begin{table*}
\begin{minipage}{17.7cm}
\begin{minipage}{5.85cm}
\caption[]{Small part of the resulting photometric data showing five
entries around G$\,$359 (see \#5 in Table~\ref{t_lmc4cskmacs}).
Stars are identified by their field sequence numbers.
The coordinates $(\alpha,\delta)$ derived from the MACS reference grid are
given along with the calibrated $V$ magnitudes, \mbox{$B-V$} colours and
the photometric errors (from DAOPHOT).
The entire table with 46$\,$749~entries is available electronically, see the
footnote to Sect.~\ref{s_lmc4c_obsred}
}
\label{t_lmc4cphot}
\end{minipage} \hspace*{0.15cm}
\begin{minipage}{11.3cm}
{\small \begin{tabular}{cr@{ }r@{ }r@{}lr@{ }r@{ }r@{}lcccc}
\hline
\noalign{\smallskip}
Field & \multicolumn{4}{c}{$\alpha$} & \multicolumn{4}{c}{$\delta$} &
 $V$ & $\Delta V$ & $B-V$ & $\Delta (B-V)$ \\
sequ. & [$^{\rm h}$ & $^{\rm m}$ & $^{\rm s}$ &] & [$^\circ$ & $'$ & $''$ &] &
 [mag] & [mag] & [mag] & [mag] \\
\noalign{\smallskip}
\hline
\tabvd & \multicolumn{4}{c}{\tabvd} & \multicolumn{4}{c}{\tabvd} & \tabvd &
 \tabvd & \tabvd & \tabvd \\
 1.010774 & 5 & 31 & 7&.233 & $-66$ & 39 & 27&.97 &
   20.728 & 0.047 & 0.818 & 0.100 \\
 1.010460 & 5 & 33 & 8&.814 & $-66$ & 39 & 27&.99 &
   21.046 & 0.048 & 0.287 & 0.068 \\
 1.112922 & 5 & 32 & 3&.793 & $-66$ & 39 & 27&.99 &
   12.543 & 0.013 & 0.189 & 0.022 \\
 3.026160 & 5 & 30 & 11&.500 & $-66$ & 39 & 27&.99 &
   18.807 & 0.060 & 0.026 & 0.102 \\
 3.025976 & 5 & 31 & 20&.923 & $-66$ & 39 & 28&.02 &
   21.388 & 0.041 & 0.473 & 0.187 \\
\tabvd & \multicolumn{4}{c}{\tabvd} & \multicolumn{4}{c}{\tabvd} & \tabvd &
 \tabvd & \tabvd & \tabvd \\
\hline
\end{tabular}}
\end{minipage}
\end{minipage}\vspace*{-0.25cm}
\end{table*}

Table~\ref{t_lmc4cphoterr} lists photometric errors derived by DAOPHOT and
their standard deviation for three magnitude intervals (comparable to the
statistical errors given in Table~2 of Braun et al. 1997).
The values for the single stars can be found in the resulting data table
(Table~\ref{t_lmc4cphot}).
The quality of all three large photometric studies inside LMC$\,$4 can be
checked by plotting the magnitude and colour differences of stars in common
(see Fig.~\ref{f_lmc4c2j}).
The mean value, standard deviation, minimum and maximum value of
the differences between the photometries of Braun et al. (1997, 'J'),
Dolphin \& Hunter (1998, 'DH') and the present one ('C') are given
in Table~\ref{t_lmc4c2j}.
Because the data were paired by software, in particular at the faint end
mismatches occured, which affect these values.
To determine better statistics, we applied selection criteria given
in the footnote of Table~\ref{t_lmc4c2j}.

Even though the new dataset covers a larger area
($A_{\rm C_{1-3}} \approx 1.1 \cdot A_{\rm J}$), the colour-magnitude diagram
of the new data (Fig.~\ref{f_lmc4ccmds}a) shows less upper main sequence stars
than the CMD of the earlier data (Fig.~\ref{f_lmc4ccmds}b).
This is caused by the larger stellar density in the E-W portion, i.e. fields
0-9 of the 'J'-shaped region containing a part of LH$\,$77.
The combined datasets cover $574\;\sq '$ on the sky, i.e. about
11\% of the LMC$\,$4 area as defined by the H$\alpha$ filaments.

To calculate absolute magnitudes we used the distance modulus of
$(m-M)_0 = 18.5\;$mag, which is still the {\it best} applicable value.
Gibson (2000) gives a range of current determinations of $[18.07,18.74]\;$mag
(see also Groenewegen \& Oudmaijer 2000).

The isochrones of the Geneva group (Schaerer et al. 1993) for LMC metallicity
($Z = 0.008$ or [Fe/H]$=-0.34\;$dex) had been fitted by eye,
yielding age and interstellar extinction as given in the next section.

\section{Results of photometric analyses}
\label{s_photdat}

To determine the age of stellar populations inside LMC$\,$4 we had already
obtained CCD photometry in $B,V$ passbands (Braun et al. 1997).
The two strips with a total of 25 CCD positions (see Fig.~\ref{f_lmc4c_hw})
cover 298~arcmin$^2$ equivalent to 6\% of the area inside the H$\alpha$ filaments.
This 'J'-shaped area constitutes an E-W strip of $400\;$pc in the region
of the OB superassociation LH$\,$77 and a S-N strip of $850\;$pc.

The colour-magnitude diagrams (CMDs) and luminosity functions (LFs) derived from
the data yielded ages of $9-16\;$Myr, a colour excess of mostly only 0.11~mag
(in some places even less), and LF slopes of 0.22-0.41 (Braun et al. 1997;
Braun 1996).
The age range given is maximized but nevertheless by a factor 2 smaller
than the necessary age spread required by the global SSPSF scenario (see
Sect.~\ref{ss_sspsf}).
Additionally, no correlation with the distance to the centre of LMC$\,$4
was found.
In all CMDs a young population of about 11~Myr is present, even though
the stellar density of this population gets low towards the north leading to
the higher LF slopes (compared to an expected slope $\gamma = 0.27$, see
Will 1996).

Dolphin \& Hunter (1998) analyzed $UBV$ photometry of four~fields inside
LMC$\,$4 and one~field outside, near its rim.
This study yields ages of 12-16~Myr and $B-V$ excess reaching from 0.04 to
0.07~mag inside LMC$\,$4 and $\sim 7\;$Myr and 0.09~mag for the NGC$\,$1955
field (i.e. the N$\,$51 region).
The lack of an age gradient inside the SGS is supported.

These studies can now be extended by the present $B,V$ CCD photometry for three
positions of the central region inside SGS LMC$\,$4
(see Fig.~\ref{f_lmc4c_hw}), which are centered near the clusters
HS$\,$343 and KMHK$\,$1000.
The observed area of 69$\,$500~pc$^2$
would contain the stellar population predicted to be 30~Myr of age
according to the model of the triggered star cluster arcs (see
Sect.~\ref{ss_arcscen}), even if that population were spatially dispersed.

The isochrone fit to the new dataset yields an age of $11\,(2)$~Myr and
a colour excess $E_{B-V} = 0.10\,(3)\;$mag.
This value was also checked by deriving intrinsic colours and thus reddening
by the Wesenheit function (see e.g. Hill et al. 1994):
$\;W := V - 3.1 \cdot (B - V)\;$
and an appropriate fit to the stellar evolution model.
We used the main sequence (MS) stars with $V \in [14.7 , 17.5]\;$mag and 
an isochrone of $\log (t/[{\rm yr}]) = 7.05$ yielding
$(B-V)_0 \approx -0.693\;\mbox{mag}  + 0.025 \cdot W$.
The $E_{B-V}$-distribution of these 629 MS stars has its maximum
at 0.1~mag with a mean (and deviation) of $0.12\,(4)\;$mag.

\begin{table}
\caption[]{Age ($t$) and reddening ($E_{B-V}$) of star populations inside
and at the edge of LMC$\,$4 sorted from north to south, see
Figs.~\ref{f_lmc4c_hw} and \ref{f_lmcsgslhsketch}}
\label{t_lmc4clitdat}
\begin{tabular}{lrrl}
\hline
\noalign{\smallskip}
Object & \multicolumn{1}{c}{$t$} & \multicolumn{1}{c}{$E_{B-V}$} &
 \multicolumn{1}{c}{Paper}\\
 & \multicolumn{1}{c}{[Myr]} & \multicolumn{1}{c}{[mag]} &\\
\noalign{\smallskip}
\hline
\noalign{\smallskip}
\multicolumn{4}{l}{\it Inside LMC$\,$4:}\\
 'C' region & 11 & 0.10 & this paper \\
 'J' region & 9-16 & $\la 0.11$ & Braun et al. 1997 \\
 4 inside fields & 12-16 & 0.04-0.07 & Dolphin \& Hunter 1998 \\
\multicolumn{4}{l}{\it Edge of LMC$\,$4:}\\
 NGC$\,$2030 & 4 & 0.19 & Laval et al. 1986 \\
 NGC$\,$1948 & 5--10 & 0.20 & Will et al. 1996 \\
 LH$\,$91, 95 & $\sim 8$ & 0.17 & Gouliermis et al. 2000 \\
 LH$\,$72, north & 8--15 & 0.00--0.04 & Olsen et al. 1997 \\
 LH$\,$72, south & 5 & 0.06--0.17 & Olsen et al. 1997 \\
 NGC$\,$2004 & 16 & 0.09 & Sagar \& Richtler 1991 \\
 LH$\,$63 & 14 & 0.07 & Petr 1994 \\
 LH$\,$60 & 9 & 0.04 & Petr 1994 \\
 NGC$\,$1955 & 7 & 0.09 & Dolphin \& Hunter 1998 \\
 LH$\,$54 & 6 & 0.10 & Petr 1994 \\
 LH$\,$76 & 2--5 & 0.09 & Wilcots et al. 1996 \\
\noalign{\smallskip}
\hline
\end{tabular}
\end{table}

The similarity of the morphology visible in the CMDs of the two datasets,
the one (Fig.~\ref{f_lmc4ccmds}a,  fields C$_{1-3}$)
containing the populations at the LMC$\,$4 centre and the other
(Fig.~\ref{f_lmc4ccmds}b, fields $0-10$ and $17-24$)
being dominated by the superassociation LH$\,$77, is striking.
No population of age 30~Myr is present in the central part.
This follows directly from the comparison of the region in both CMDs
to the right of the main sequence, near the $30\;$Myr isochrone.

The members of the group of six A-type supergiants near HS$\,$343
(see Fig.~\ref{f_lmc4arcs}; these stars selected by Efremov \& Elmegreen 1998
from the data of Rousseau et al. 1978 are No. 5, 8, 9, 10, 12, and 13
of Table~\ref{t_lmc4cskmacs}), quoted as relicts of a $30\;$Myr population,
are marked by crosses in Fig.~\ref{f_lmc4ccmds}a.
The CMDs for the two clusters near the geometric centre of the filamentary
H$\alpha$ boundary show their ages to be $\sim 0.1\;$Gyr for HS$\,$343 and
$\sim 0.3\;$Gyr for KMHK$\,$1000 (see Fig.~\ref{f_lmc4cCLcmds}).

\section{Age structure inside LMC$\,$4}
\label{s_agestruc}

The similarity of the morphology visible in the CMDs of all datasets 
is striking.
Therefore the stars inside LMC$\,$4 must have been formed in a short period
of time,
so the H$\alpha$ and H$\,${\sc i} feature called LMC$\,$4 must have been
created in a turbulent manner rather than in a 'bubble blowing' action.
This fits also to the small (i.e. $\ll 500\;$pc) disk thickness of the LMC
(e.g. Kim 1998 derived in Sect.~5.2.1 an exponential scale height of the gas
of \mbox{$\sim 170\;$pc}, i.e. a disk thickness of $\sim 340\;$pc).
This implies that this SGS resembles a truncated cylinder rather than a bubble.
The onset of this star burst can be dated by isochrone fitting
giving as average age of the interior population $12\,(2)$~Myr.
No other population of age less than 100~Myr is present in the central part.

The star groups near the edge are younger indeed than the stars in the 
interior of LMC$\,$4. 
The age at the edge (age of stars associated with the H$\alpha$ emission
outside the H$\alpha$ rim of the SGS) is 4-7~Myr
(see Table~\ref{t_lmc4clitdat} and Sect.~\ref{ss_evoaftbrst}).
This implies that $\sim 5\;$Myr after the giant central volume formed stars,
the edge was pushed into star formation mode, too. 

One is thus faced with several questions pertaining to the origin of 
the structure called LMC$\,$4.
What mechanism can stimulate {\it simultaneous} star formation over an area 
(in a volume) with projected size of about $800\mbox{ pc} \times 1200\;$pc 
(the interior of LMC$\,$4)? 
By what mechanism was the gas of the birth cloud dispersed 
so efficiently to create the presently low column densities 
in the H\,{\sc i} over such an area?
What stimulated the star formation at the outskirts of this area 
(creating the H$\alpha$ structure called LMC$\,4$) 
in a relatively homogeneous manner? 

\section{Discussion of scenarios proposed for LMC$\,$4's creation}
\label{s_models}

Several star formation scenarios possibly leading to the formation of
supergiant shells are mentioned in the literature.
Here we will comment on each in relation with our knowledge of LMC$\,$4.

\subsection{HVC infall}
\label{ss_hvc}

{\sf An infalling high-velocity cloud might compress the gas in 
the LMC disk and so trigger star formation.}

Infall would affect star formation in a large region indeed and we see
that the ages derived for the large inner part of LMC$\,$4 are equal
(see Sect.~\ref{s_photdat}).
However, the data about the density and radial velocity of LMC$\,$4 gas
contradict this scenario (Domg\"orgen et al. 1995),
and infall was not the actual trigger for the formation of LMC$\,$4.

In addition, given the motion of the LMC through the halo of the Milky Way
(Kroupa \& Bastian 1997),
the velocity of such a HVC should have been rather orthogonal
to that of the LMC itself,
thus having a velocity vector pointing more or less toward the Sun.
A HVC moving away from the Milky Way is very unlikely.
HVCs moving toward the Milky Way at $+60$ and $+120\;{\rm km\;s}^{-1}$
are known in this direction (Savage \& de Boer 1979).
Since this high velocity gas is seen over the entire face of the LMC
(see Savage \& de Boer 1981),
with the LMC at $v_{\rm rad} \approx 250\;{\rm km\;s}^{-1}$,
it is rather nearby. 

\subsection{SSPSF}
\label{ss_sspsf}

{\sf In the stochastic self-propagating star formation scenario 
the stars formed initially trigger further star formation in their vicinity.}

SSPSF leads to an age gradient from the starting point, 
e.g. the centre of a SGS, outward to the rim (Feitzinger et al. 1981).
It has been a popular model and was often cited in papers about
supershells in distant galaxies 
where no photometry of individual stars is available.

The predictions by SSPSF for the morphology of SGSs agree with the observed
density and radial velocity of gas in the interior of LMC$\,$4.
However, the large radius of the SGS would lead to a continuous age spread
of the stars of at least 15~Myr range.
Even if one takes a possible overlap of stellar populations having different
ages into account, 
the photometric data are in contradiction with such a population structure
and thus indicate incompatibility of the SSPSF scenario with the interior
of LMC$\,$4.
This has now been sufficiently noted 
(see Braun et al. 1997; Dolphin \& Hunter 1998).

\subsection{Triggering of stellar arcs}
\label{ss_arcscen}

{\sf Star clusters or $\gamma$-ray bursts may trigger star formation 
to form arc like structures.}
 
\begin{figure}
\epsfxsize=8.4cm
\centerline{\epsffile{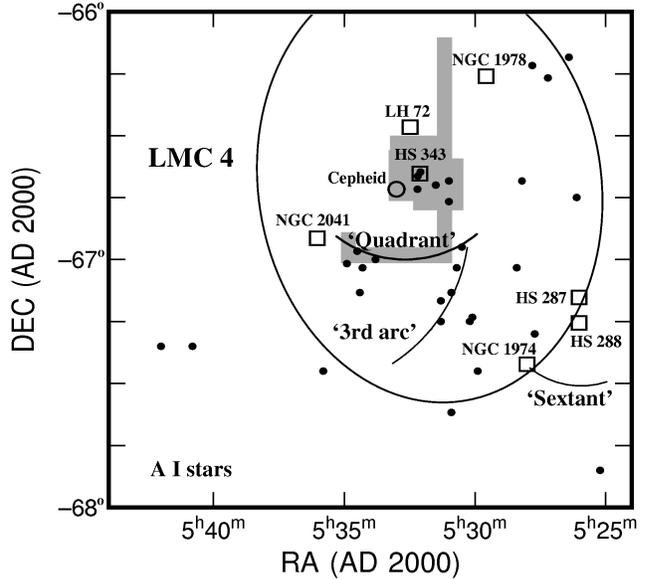}}
\caption[]{Schematic plot of selected structures near LMC$\,$4.
The ellipse shows the location of the H$\alpha$ filaments and
the H$\,${\sc ii} regions, the gray area the regions covered by the
present and the earlier (Braun et al. 1997) photometry.
Some of the structures mentioned by Efremov~\& Elmegreen (1998)
in relation with arc formation (the arcs, some AI stars from
Rousseau et al. 1978, the Cepheid HV$\,$2667, erroneously given as
HV$\;$5924 by Efremov \& Elmegreen)
are indicated (figure adapted from Fig.~2 of Efremov \& Elmegreen 1998)
}
\label{f_lmc4arcs}
\end{figure}

Efremov \& Elmegreen (1998) suggested as trigger for star formation
in LMC$\,$4 the self-gravitational collapse of parts of rims from
shells initially driven by star clusters,
leading to the building of star cluster arcs.
They identify three of these structures (see Fig.~\ref{f_lmc4arcs}), 
circular (in the 2d projection) in shape,
and this may indicate that they have been formed by swept-up material.
Efremov et al. (1998) subsequently suggested 
that gamma-ray bursts (GRBs) may produce the pressurizing wave. 

These suggestions are rather {\it ad hoc}. 
Moreover, they would create localized structures 
and nothing of the kind like LMC$\,$4. 
The collapse into a cluster or the explosion of a GRB, 
formation of a shell of swept up material which itself forms stars, 
which then leads to a next shell or the acceleration of the old shell 
(the scenario for the `Quadrant' and subsequently for LMC$\,$4 after
Efremov \& Elmegreen 1998) 
may explain some features, but not the entire SGS with the documented
age structure (see Sect.~\ref{s_agestruc}).
Furthermore it is hard to explain why `Quadrant' is coeval (Braun et al. 1997)
while `Sextant' shows a clear age gradient (Petr et al. 1994) if the same
mechanism would be valid.
The geometry of the (after Efremov \& Fargion 2000) four cluster arcs
are rather due to cloud structure long {\it before} star formation than
being directly connected to the formation process.

\begin{figure}
\epsfxsize=8.4cm
\centerline{\epsffile{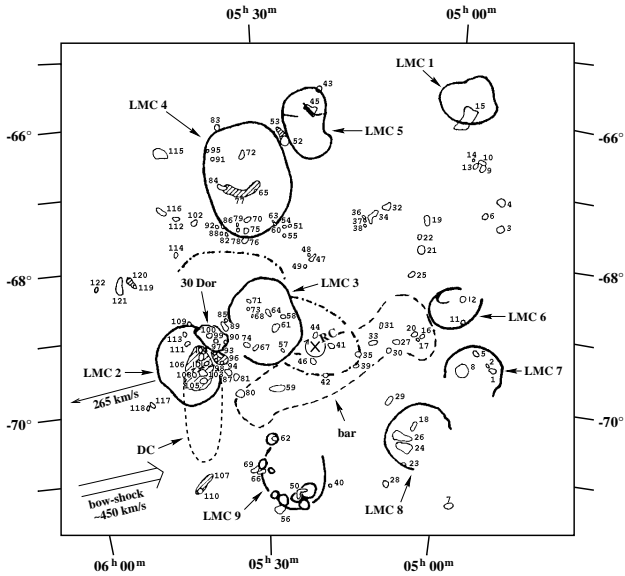}}
\caption[]{Sketch of the LMC with the nine supergiant shells
(LMC$\,$1 - LMC$\,$9) and the giant H$\,${\sc ii} region 30$\,$Doradus
(from Meaburn 1980, Fig.~2),
the dark cloud (DC) south of 30$\,$Dor,
the 122 OB associations (LH$\;nn$, from Lucke \& Hodge 1970, Fig.~1),
and the isophote of the bar (from Smith et al. 1987, Fig.~4).
The LMC has an imprecisely defined rotation centre, situated near RC.
The movement of the LMC in the galactic halo is indicated by
the direction of the proper motion.
The resulting bow-shock, driven by the sum of motion and rotation,
is $\sim 450\;$km s$^{-1}$ (see de Boer et al. 1998)
}
\label{f_lmcsgslhsketch}
\end{figure}

\subsection{Star formation triggered by the LMC bow-shock}
\label{ss_bowscen}

{\sf The motion of the LMC through the Milky Way halo leads to a 
bow-shock triggering star formation.} 

Since a very large area created stars at the same time, 
and since (due of the rotation of the LMC) 
the present day LMC$\,$4 was near the leading edge about 15~Myr ago, 
the bow-shock of the LMC may have triggered the conditions for
large-scale star formation. 
Searching for a possible creation mechanism of supergiant shell LMC$\,$4
and maybe of all the supershells of the Large Magellanic Cloud,
one finds that the superstructures of the LMC 
(see Fig.~\ref{f_lmcsgslhsketch}) 
are all located at the outskirts. 
Looking at these features clockwise from south-east to north-west,
the expected correlation of age with traversed distance is visible 
(see de Boer et al. 1998).

The scenario has as essential part the motion of the LMC through the galactic
halo and the LMCs clockwise rotation.
The gas of the LMC rotates through the compression zone 
at the south-east, to eventually find itself in the north-east (LMC$\,$4) 
with all the consequences of the star formation 
triggered inside that gas.

\section{Summary and conclusions}
\label{s_sumcon}

In the review of the scenarios above we have indicated
that the first three are not compatible with the observational facts.
The most important fact is that over the entire interior of LMC$\,$4
star formation burst out on a large scale about $12\;$Myr ago
(see Table~\ref{t_lmc4clitdat} and Sects.~\ref{s_photdat} and \ref{s_agestruc}).
Small stellar associations cannot be used as trigger for
large-scale structures
and there is no evidence for a recent collision with a HVC.
The bow-shock formation scenario is the only viable one.
Note that the bow-shock scenario explains only
the formation of large structures around the edge of the LMC.
At any given location in the LMC star formation continues as is
readily visible from the ubiquitous H$\alpha$ structures
everywhere in the LMC.

\subsection{Evolution after the star burst took place}
\label{ss_evoaftbrst}

After formation of the original cloud of stars
these stars emitted their energy into the surroundings,
driving the gas out of this region
rather more in a turbulent than a pressurized manner.
The total energy produced by the massive stars,
(stellar winds, ionizing photons, and supernova explosions)
went into ionizing and expansion of the original birth cloud.
These aspects have been discussed by Braun et al. (1997) and by
Dolphin \& Hunter (1998) and are not repeated here.

After the formation of this 1st generation of stars,
star formation was initiated at the edge of the central region,
creating populations of younger ages
visible by the H$\,${\sc ii} regions surrounding this SGS.
Examples for these stellar populations are LH$\,$63, LH$\,$60 and LH$\,$54
(Petr et al. 1994), situated in the N$\,$51 region,
and LH$\,$76 (Wilcots et al. 1996) more to the east
(see Fig.~\ref{f_lmcsgslhsketch} and Table~\ref{t_lmc4clitdat}).
Also part of LH$\,$72 (Olsen et al. 1997) is one of these,
assuming it lies somewhat outside the plane of LMC$\,$4,
being at a different depth in the LMC.
This phase with the formation of stars at the outskirts of the original cloud
affects only regions of small size (i.e. $< 300\;$pc)
as expected from the limits due to inhomogenity
(density, temperature, magnetic fields etc.) of the medium.

\subsection{Details of the bow-shock formation and
consequences for the overall structure of the LMC}
\label{ss_bsdet}

The convergence of gas near the bow-shock,
due to the LMC space motion combined with the LMCs rotation
(see de Boer et al. 1998)
creates a large and condensing gas cloud.
In it star formation sets in, and the rotation of the LMC
moves this region around the edge in clock-wise fashion.

The role of ram-pressure, especially the comparison of strength of
hydrodynamic and gravitational effects, is still uncertain.
Observational findings like X-ray shadowing (Blondiau et al. 1997),
the steep edge in H$\,${\sc i} near the location of the largest velocity
with respect to the halo gas (see e.g. the map by Kim et al. 1999),
and H$\alpha$ emission of some main components of the Magellanic Stream
(Weiner \& Williams 1996) are all indications for bow-shocks
due to motion of Magellanic material through halo gas.
To fit the H$\,${\sc i} maps of the Stream and other structures,
recent models for the tidal interactions of the Magellanic Clouds (MCs)
are extended by a drag term (Gardiner 1999), accounting for the motion
of the MCs through the Milky Way halo.
However, problems remain and not all inconsistencies are solved
(see e.g. Moore \& Davis 1994 and Murali 2000).

The shape of the dark cloud presently seen south of 30 Dor is elongated,
oriented perpendicular to the direction of the rotation centre of
the LMC.
The most dense region of stars inside LMC$\,$4,
being the hatched region in Fig.~\ref{f_lmcsgslhsketch}
with the embedded smaller associations LH$\,$65 and LH$\,$84,
has the same orientation.
We note that at their radial distance from the centre
(both at approximately 2~kpc)
the LMC rotation curve is rather level,
and rotation will hardly change the original shape of structures.

Since the density of the LMC drops radially outward,
it is to be expected that older SGSs will have opened up to the LMC edge.
The morphology of the H$\alpha$ distribution may therefore give
further evidence for the expected age sequence of the SGSs.
Continuing in clockwise direction, the supergiant shell LMC$\,$5
to the east of LMC$\,$4 is still complete,
while LMC$\,$6 is a non-contiguous ring.
Note also, that the star density in LMC$\,$4 drops outward and
that LMC$\,$4 is elongated toward the lower density on the outside.

Apart from the young stars in LMC$\,$4 older populations are present.
These may date from a previous burst, of one or more revolutions ago.
Such earlier events will have left numerous earlier generations.
However, a bow-shock triggered burst need not take place continuously,
since apparently gas accumulates intermittantly and the power of the trigger
is sensitive to the current halo density at the front of the LMC.
If it were continous,
one would see a continuous ring of young stars around the LMC's edge.

The age of small-scale structures may or may not be related
to the age of the large-scale structures such as LMC$\,$4.
One full rotation of the LMC edge takes $\simeq 100\;$Myr
so that larger older structures may be part and/or consequence
of a previous large-scale event.

\subsection{Comparison with other dwarf galaxies}
\label{ss_dgcomp}

Looking at other dwarf galaxies in H$\alpha$ or H$\,${\sc i}, one recognizes
that supershells are quite common features in galaxies and especially in dwarf
galaxies (e.g. Brinks \& Walter 1998),
where the symmetry is not disturbed by differential rotation.
However the different environment of these galaxies,
such as tidal forces from neighbouring galaxies or being completely isolated,
makes it difficult to explain these supershells by the same creation mechanism
as the supergiant shells in the Magellanic Clouds.
Still, in such galaxies larger volumes of space must attract
sufficient amounts of gas in dense form that a starburst takes place,
setting off star formation nearly simultaneous in huge volumes.

Clearly, photometric studies of resolved stellar populations
in more distant (dwarf) galaxies are needed
to achieve a general picture on these large-scale star forming processes.

\subsection{Conclusions}
\label{ss_concl}

The photometry of large areas inside LMC$\,$4 shows that the stars
are of the same age, being $\simeq 12\;$Myr.
The rotation speed of the LMC makes clear that the LMC$\,$4 region was
at time 0 near the bow-shock.
At the bow-shock one currently finds a huge dark cloud
being in the process of forming stars.
Combining these two observational facts led to the bow-shock scenario
and further evidence has been brought forward to substantiate this scenario.

Observing the interior of further supergiant shells
will provide clearer insight in the formation history
of the star groups inside such structures.
Except for the complicated 30$\,$Dor region
(with the neighbouring supergiant shells LMC$\,$2 and LMC$\,$3)
all other SGSs should show a coeval large-scale stellar population
fitting to the age sequence proposed by the bow-shock scenario.
Were this found to be not so,
then the bow-shock trigger model would be falsified.
One then would be in need of finding another star formation trigger.
Given the current observations the only other possible
(however unlikely) scenario would then be a special HVC collision.
As yet, the bow-shock induced star formation fits the available
observational data best.

\section*{Acknowledgments}

JMB was partly supported through the Deut\-sche For\-schungs\-ge\-mein\-schaft
(DFG) in the {\it Graduiertenkolleg} `The Magellanic Clouds and other Dwarf
Galaxies', MA by the DFG through grant Bo 779/21.

We wish to thank Dr. Deidre A. Hunter for providing us a CCD image of
their LH$\,$77 field for star identification.

\end{document}